\documentclass{aa}

\usepackage{txfonts}
\usepackage{graphicx}
\usepackage{epsf}

\begin{document}

\title{Gas mass fraction from \emph{XMM-Newton} and \emph{Chandra} high redshift clusters and its use as a cosmological test}

\titlerunning{Gas mass fraction of distant clusters and its use as a cosmological test}

\author{
L.D. Ferramacho
\and 
A. Blanchard
}

\institute{Laboratoire d'Astrophysique de Tarbes et Toulouse, OMP, CNRS, UMR 5572, UPS, 14, Avenue E. Belin, F-31400 Toulouse, France\\
\email{luis.ferramacho@ast.obs-mip.fr \& alain.blanchard@ast.obs-mip.fr}}

\offprints{luis.ferramacho@ast.obs-mip.fr}

\date {Received / Accepted}

\abstract
   {}
   {We investigate the cosmological test based on the evolution of the gas fraction in X-ray galaxy clusters and the stability of the cosmological parameters derived from it.}
   { Using a sample of distant clusters observed by \emph{XMM-Newton} and \emph{Chandra}, and comparing their gas fraction at different radii to the gas fraction observed for nearby clusters, we have determined the likelihood functions for $\Omega_m$ in a flat universe and the confidence contours in the $\Omega_m-\Omega_{\Lambda}$ plane.}
   { Results obtained at the virial radius point to a high matter density Universe, while for inner radii the $\Omega_m$ parameter obtained tends to decrease, reaching values compatible with the concordance model. The analysis allows us to conclude that this test provides ambiguous results due to the complex structure of the ICM that induces a dependence of the gas fraction on temperature, radius, and redshift, which cannot be accounted for by the self-similar picture expected from pure gravitational heating of the ICM.}
   {The use of gas fraction in X-ray clusters to constrain cosmological parameters seems therefore to be compromised until a better understanding of the ICM physics and the ability to obtain observations of better quality up to the virial radius are achieved.}


\maketitle

\section{Introduction}

Physical proprieties of galaxy clusters have been considered a promising tool for cosmological probes.
A test based on the assumption of a constant baryon fraction in these objects was proposed by \cite{sasaki} and \cite{pen}.
These authors showed that under this assumption, the gas fraction of clusters can, in principle, be used as a standard ``ruler'' to determine
cosmological parameters. The main idea is that if the only heating source of the gas is gravitational collapse, the baryon fraction should be
constant and independent of redshift. Since the measured baryon fractions
depend on the cosmology through the angular distance to the clusters, the ``true'' cosmological model would be one where all the measured baryon fractions at different
redshifts would be equal.   
This method has been applied in the past years (\cite{Rines}; \cite{ettori1}; \cite{Allen2002}; \cite{Ettori}; \cite{Allen2004}) using 
clusters observed by \emph{Chandra} and \emph{ROSAT PSPC}, with the results pointing to a low value of $\Omega_{m}$, consistent with the concordance model ($\Omega_m\approx0.3,\Omega_{\Lambda}\approx0.7$). 
However, these authors have often restricted the method to a single radius, defined by a given high overdensity relative to the critical density of the Universe.
To check the validity of this cosmological test, it is desirable to perform a good study on the behavior of the gas fraction distribution (the main contribution of baryons) inside clusters and its evolution with redshift. Such a study was done recently in \cite{sadat05}, hereafter S05, who claim that clusters are more complex objects than it was expected in self-similar models, as they found that gas fraction presents a dependence on radius, temperature, and redshift.

The goal of the present work is to give a quantitative analysis on the results obtained in S05, by determining the likelihood functions of the cosmological test based on the gas fraction for the sample of S05 and study the effect of using radii of different overdensities to perform it. The paper is organized as follows. In Sect. 2, we present the cluster samples that was used in the test. In Sect. 3, we present the method used to determine the gas fractions for all clusters in the sample. The cosmological results are presented in Sect. 4 and finally, in Sect. 5, we draw our conclusions. We used a Hubble constant of 50 km/s/Mpc, unless the dependence is explicitly given (with $H_0$=100$h$km/s/Mpc).        

\section{The cluster sample}

We used three samples of clusters taken from different references. The nearby sample consisted of 35 nearby clusters with redshifts between $0.02$ and $0.095$, observed by \emph{ROSAT PSPC} and compiled in \cite{vfj99}. This sample contains symmetric clusters with high-quality imaging data up to large radii, which allows for good estimations of the gas masses on the outer parts of the clusters.
We have also used XMM data obtained on a sample of eight high-redshift clusters observed as part of the \emph{XMM-Newton} $\Omega$ project, a systematic $XMM-Newton$ guaranteed time follow-up of the most distant SHARC clusters \cite{bartlett}. The high sensitivity of $XMM-Newton$ allows for an investigation of the emissivity in high-redshift clusters beyond half the viral radius, a remarkable result \cite{Arnaud}. The sample contains clusters within a restricted range of redshifts, between $0.45$ and $0.65$, with the detailed data reduction and analysis being presented in \cite{lumb}.

To have a bigger statistical set of high-redshift clusters, we considered another sample of eight clusters observed by \emph{CHANDRA}, within the same range of redshifts ($0.451$ to $0.583$) as the above XMM clusters. This sample was taken from \cite{vik2002}, who claim to have an accurate measurement of the X-ray brightness nearly up to the virial radius.

For all clusters, we had information about the temperature, surface brightness parameter $\beta$ and core radius r$_c$ taken within the observed region of the clusters, and some information (depending on the sample) that allowed us to normalize the gas mass profile, all the ingredients needed to determine the gas mass fraction as it will be shown in the next section.  

\section{Gas mass fractions}    

\subsection{Determination of the gas mass }

The gas mass fraction of a cluster at a given radius $r$ is defined as $f_{gas}(r)=M_{gas}(r)/M_{tot}(r)$, needing good estimations of the gas mass and the total mass contained inside a radius $r$. The gas mass is generally determined by integration of the so-called $\beta$-model for the description of the X-ray gas profile \cite{Cavaliere}:
\begin{equation}
\rho_{gas}(r)=\rho_{gas}(0)\left[1+\left(\frac{r}{r_c}\right)^2\right]^{-\frac{3}{2}\beta}\;,
\end{equation}
which translates to the observed X-ray surface brightness:
\begin{equation} 
S_X(\theta)=S_X(0)\left[1+\left(\frac{\theta}{\theta_c}\right)^2\right]^{-3\beta+\frac{1}{2}}\;.
\end{equation}
The observed brightness is due to the well-known process of bremshtralung radiation, and so the central density $\rho(0)$ can be determined from the central brightness $S_X(0)$, allowing a determination of the gas mass at any radius of detection with a good accuracy. It is important to notice that the obtained results depend on the assumed cosmology because the conversion of the observed angles to physical radii depends on the angular distance to the cluster as $r=\theta d_A(z;\Omega_m,\Omega_{\Lambda})$. Using the $\beta$-model we were then able to determine the gas mass profiles of our sample of clusters within the range of detection, for any given set of cosmological parameters. When necessary, the gas mass beyond the radius of detection and up to the virial radius was estimated by extrapolation using a $\beta$-model with parameters obtained within the region of detection.        

\subsection{Determination of the total mass}

In the simplest picture of purely gravitationally driven formation of virialized systems such as galaxy clusters, it is expected that such objects exhibit self-similarity (Kaiser 1986), meaning that the radial profile of any physical quantity, such as density, should exhibit a similar shape that is independent of the cluster's reference mass and redshift, when normalized to the virial radius. Indeed, numerical simulations in which only gravitational physics is taken into account indicate that halos of different masses follow a universal density profile \cite{nfw}, the so-called NFW density profile. We considered this profile with a concentration parameter of 5:
\begin{equation}
\frac{\rho(r)}{\rho_c}=1500\frac{r^3_{200}}{r(5r+r_{200})^2}\;\;\;\;\;\;\;(c=5)\;.
\end{equation}
The normalization of this profile can be obtained using the total mass at the virial radius, which can be determined by its definition:
\begin{equation}
M_V=\frac{4}{3}\pi\bar{\rho_0}\left(1+z\right)^3\Delta_V R_V^3\;.
\end{equation} 
In this expression $\Delta_V$ is the virial contrast overdensity compared to the mean density of the universe at redshift z and $\bar{\rho_0}$ is the mean density of the Universe today. Since many authors use a contrast overdensity $\Delta_C$ compared to the critical density of the Universe ($\rho_c(z)=3H(z)^2/8\pi G$), we found it useful to make our analysis using both definitions to compare with different published results and confirm that results are independent of the adopted definition. With this second definition the last expression becomes:
\begin{equation}
M_V=\frac{4}{3}\pi\rho_c(z)\Delta_C R_V^3\;.
\end{equation} 
The virial overdensity is usually computed from the spherical top-hat model for gravitational collapse, and for an Einstein-de Sitter universe, one has $\Delta_V=\Delta_C=18\pi^2$. We used the formulae given in \cite{BN}, which provide the method to determine $\Delta_V$ and $\Delta_c$ for any flat cosmological model.   
The scaling of the mass-temperature relation can be obtained from the expressions above, using the fact that for a purely gravitational heating of the gas, $T\propto GM/R_V$, which yields:
\begin{equation}
T=A_{TM}M_{15}^{2/3}\left(\frac{\Omega_M \Delta_V}{178}\right)^{1/3} h^{2/3}(1+z)\;\rm keV
\end{equation}      
or,
\begin{equation}
T=B_{TM}M_{15}^{2/3}\left(h^2\Delta_C E^2\right)^{1/3}\; \rm keV\;,
\end{equation}    

\begin{center}
\begin{figure*}[ht]
\begin{tabular}{r l}
\includegraphics[scale=0.45]{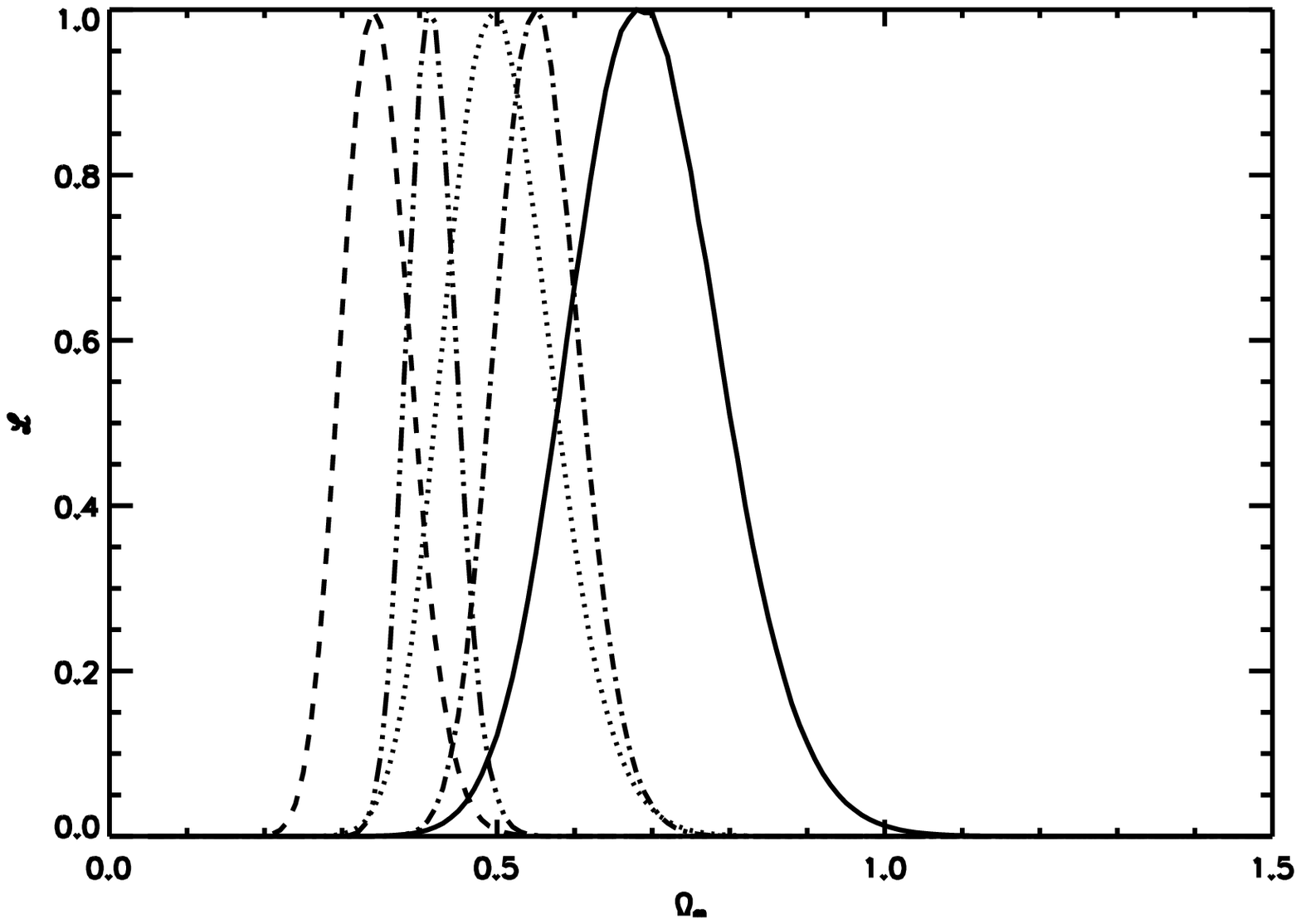}&\includegraphics[scale=0.45]{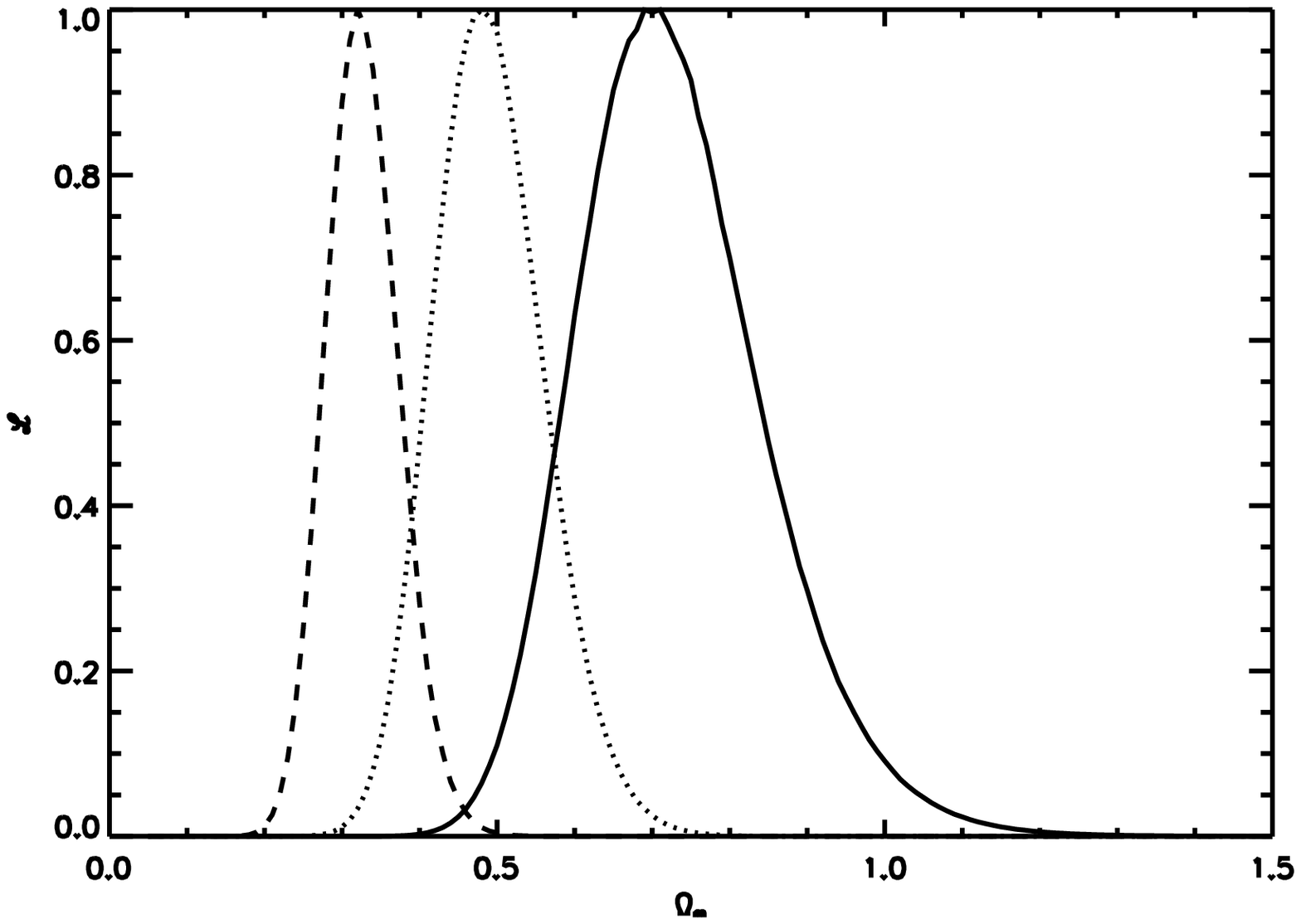}\\
\includegraphics[scale=0.45]{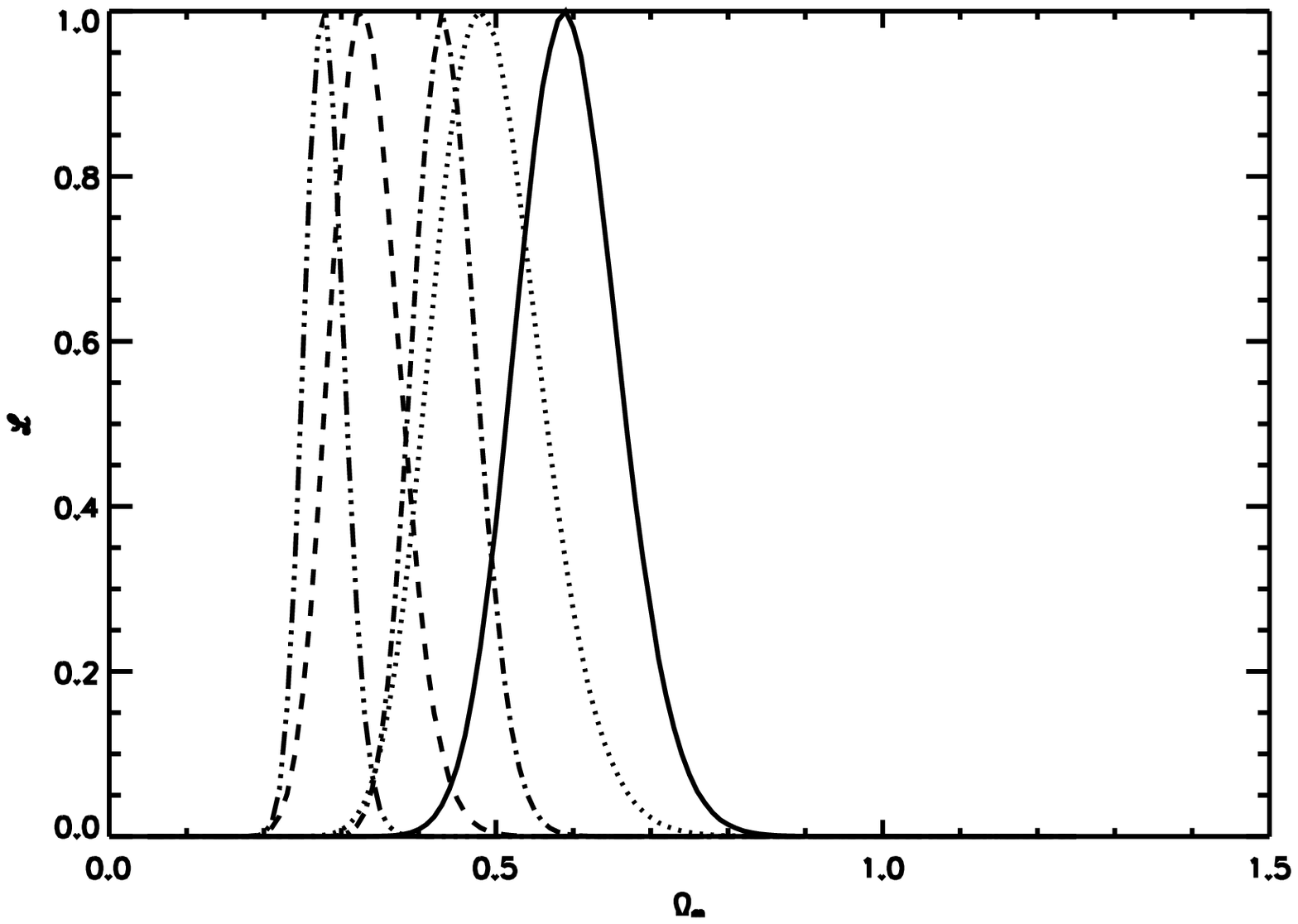}&\includegraphics[scale=0.45]{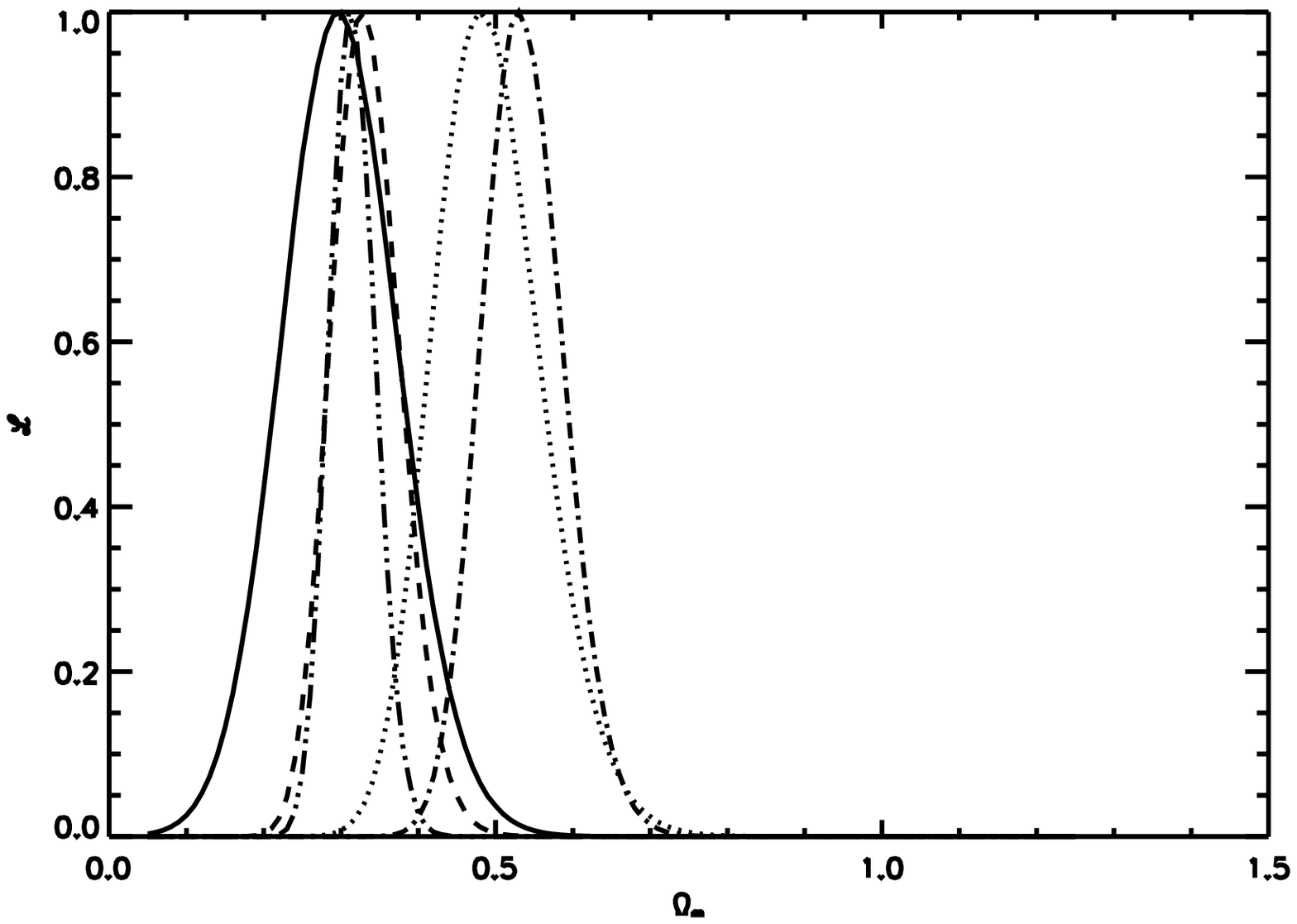}
\end{tabular}
\caption[]{Likelihood functions of the gas mass fraction test at different reference radii, for a flat universe. \emph{Left panels}: Assuming a constant normalizations $A_{TM}$ of 6.24 keV, (\cite{BN}, \emph{top}) and 11.4 keV (\emph{down}). \emph{Top right}: Assuming a normalization $A_{TM}$ that depends on $\Omega_m$ as $A_{TM}=4.9\Omega_m^{-0.75}$ KeV. \emph{Lower right}: Likelihood functions using normalization A$_{TM}$=11.4 keV, but performing the test at the same physical radii as if A$_{TM}=6.24$. Solid lines correspond to the test applied at the virial radius. The other lines correspond to likelihood functions obtained for different raddi: R$_{1}$ (dot), R$_{2}$ (dashed), R$_{500}$ (dashed-dot), and R$_{2500}$ (dashed-dot-dot).} 
\end{figure*}
\end{center}
where $M_{15}$ is the mass virial mass expressed in units of 15$M_{\odot}$ and $H(z)\equiv H_0^2 E^2=H_0^2 [\Omega_m(1+z)^3+\Omega_R(1+z)^2+\Omega_{\Lambda}]$. The normalization constants $A_{TM}$ or $B_{TM}$ can be determined from numerical simulations (\cite{evrard}; Bryan \& Norman 1998) and observations. Several studies have been performed to test the $M_V-T$ relation as predicted from numerical simulations by means of X-ray observations (see, e.g., Horner et al. 1999; Nevalainen et al. 2000; Finoguenov et al. 2001; Sanderson et al.2003; Pedersen \& Dahle 2006). Some disagreement has been found concerning the normalization constant $A_{TM}$ and the slope (steeper than the predicted 1.5) for cooler systems (T$_X$ less than 4 keV). Recently, \cite{blancharddouspis} considered an alternative way to determine $A_{TM}$. By assuming that the observed baryon content of clusters is representative of the Universe's baryon content and using the WMAP tight constraint on $\Omega_b$ and the relation between $\Omega_m$ and $h$, they derived the mass-temperature normalization as a function of $\Omega_m$, obtaining $A_{TM}=4.9\Omega_m^{-0.75}\pm 10\%$, for 0.3$\leq\Omega_m\leq$1.0.                

\section{Cosmological results}

As it was mentioned, the use of the gas fraction of clusters as a cosmological test has been based on the fact that this physical quantity should be constant at a given radius when  normalized to the virial radius. However, S05 claims that the baryon fraction of nearby and distant clusters presents a clear temperature dependence, which is larger in the inner parts and appears to be very slight at the virial radius. This fact indicates that, in principle, only the virial radius, if any, can be used as the reference radius to perform the cosmological test based on a constant gas fraction. One can relax this assumption, and consider that the dependence of the gas fraction on temperature for high redshift clusters should be related through scaling arguments to the same dependence observed for low-redshift clusters. Indeed, if the gas mass fraction varies with T, scaling would imply:
\begin{equation}
f_g\left(R/R_V,T,z\right)=f_g\left(R/R_V,T\times T_*(0)/T_*(z),z=0\right),
\label{scaling}
\end{equation}
where T$_*(z)$ is a characteristic temperature associated with a characteristic mass scale at epoch z and defined by $\sigma$(M$_*,z) \approx \delta_{th} $. So, to use the observed gas fractions for local clusters and compare them to the distant ones, we proceeded as follows: for any given set of cosmological parameters, gas fractions were determined for the local sample of clusters, and a fit was made to determine an analytical expression for its temperature dependence in the form of $f_g(R_X)=C_X\,T^{\alpha_X}$. This was done for five different normalized radii - the virial radius, two fiducial radii R$_{1}$ and R$_{2}$ (defined in \cite{vfj99} as R$_{1000}$ and R$_{2000}$, with the index referring to contrast overdensities relative to the background baryon density), and also two radii commonly used to analyze the gas fraction that correspond to more central regions of the clusters - R$_{500}$ and R$_{2500}$. For these radii, the index are referred to overdensities related to the critical density of the Universe.
The results of the fits together with Eq. (\ref{scaling}) provide a model that allows the use of high-redshift gas fractions to determine the cosmological parameters, since the ``true'' cosmological model should allow us to have $f_g^{high-z}(R_X,z,T)=C_X\,(T_*(0)/T_*(z))^{\alpha_X}\,T^{\alpha_X}$.   
The likelihood function for this test can then be written as $\mathcal{L}$=exp$(-\chi^2/2)$, with $\chi^2$ being:
\begin{equation}
\chi^2=\sum_{i=0}^N\left(\frac{f_{g,i}(\Omega_m,\Omega_{\Lambda},z_i)-f_{g}^{mod}(T_i,z_i;\Omega_m,\Omega_{\Lambda})}{\sigma_{disp}}\right)^2\;.
\end{equation}
In this expression, $N$ is the number of distant clusters in our sample and $\sigma_{disp}$ is the dispersion around the best-fit of the local cluster distribution. The errors on the derived baryon fraction for individual clusters ($\sigma_i$) were neglected since they are much smaller than $\sigma_{disp}$. 

We used three different approaches to the normalization of the mass-temperature relation: two constant values $A_{TM}=6.24$ keV (Bryan \& Norman 1998) and $A_{TM}=11.4$ keV, a $\Omega_m$-dependant value $A_{TM}=4.9\Omega_m^{-0.75}$ (Blanchard \& Douspis 2005), and finally, we assumed $A_{TM}=11.4$ keV, but taking the gas fractions at the same physical radii as if $A_{TM}=6.24$. This last approach was done in order to cross-check the dependence of the results on both the assumed radii (physical) to perform the test and the normalization constant. For the case of an $A_{TM}$ dependent on $\Omega_m$, we did not make the analysis at R$_{500}$ and R$_{2500}$ because at these radii Eq. (7) cannot be applied since we do not have an expression for $B_{TM}(\Omega_m)$.

The obtained results are presented in Fig. 1. At first glance, the main feature in each plot is the fact that the likelihood functions are highly dependent on the considered radius: $\Omega_m$ obtained from extreme radii
 are mutually exclusive and cannot therefore be combined to obtain a convergent
value for $\Omega_m$. It is important to examine whether this result depends
on the calibration of the mass-temperature relation, and we have addressed this question in some detail: in the same figure we can see the effect of using different approaches to the normalization constant. After increasing $A_{TM}$ by a factor of 2 (left panels), no significant changes are observed since the maximum likelihood values decrease only slightly. The same is observed when we used the $\Omega_m$-dependant normalization, where the results are similar to the case of $A_{TM}=6.24$ keV. 
This is consistent with the fact that changing the normalization modifies mass in a similar way for low and high redshift clusters. 
Finally, when we considered a value of 11.4 keV for $A_{TM}$, but taking the gas fractions at the same physical radii as with $A_{TM}=6.24$ keV, the likelihood functions remain almost unchanged, except for the virial radius. At this specific radius we can see a great decrease in the maximum likelihood value, since the radius is now much larger than the actual virial radius. This shows that this test is less dependent on the adopted $M-T$ normalization at inner radii, although even at this radii we cannot obtain compatible likelihoods for $\Omega_m$. 

\begin{figure}[ht]
\includegraphics[scale=0.5]{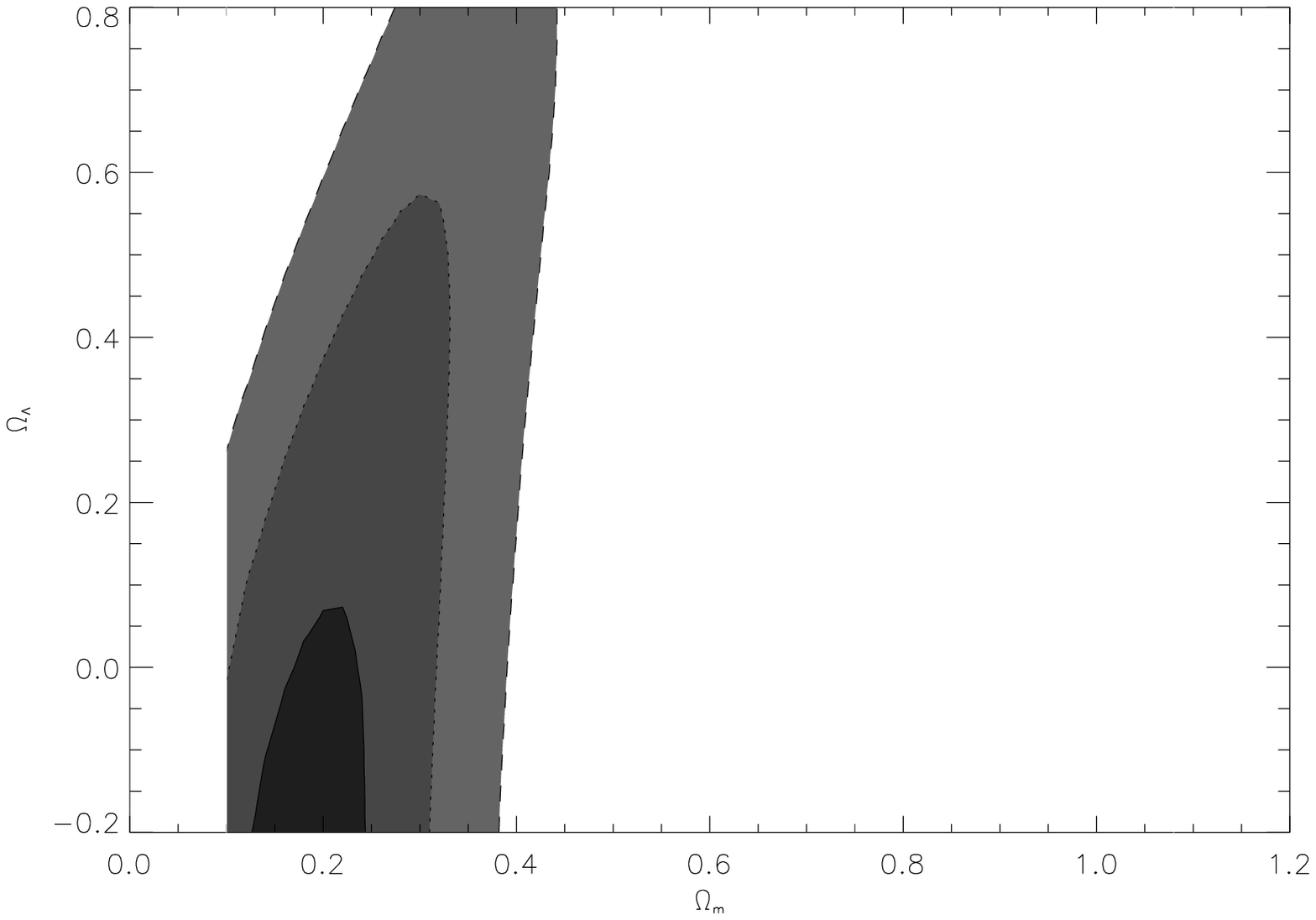}\\
\includegraphics[scale=0.5]{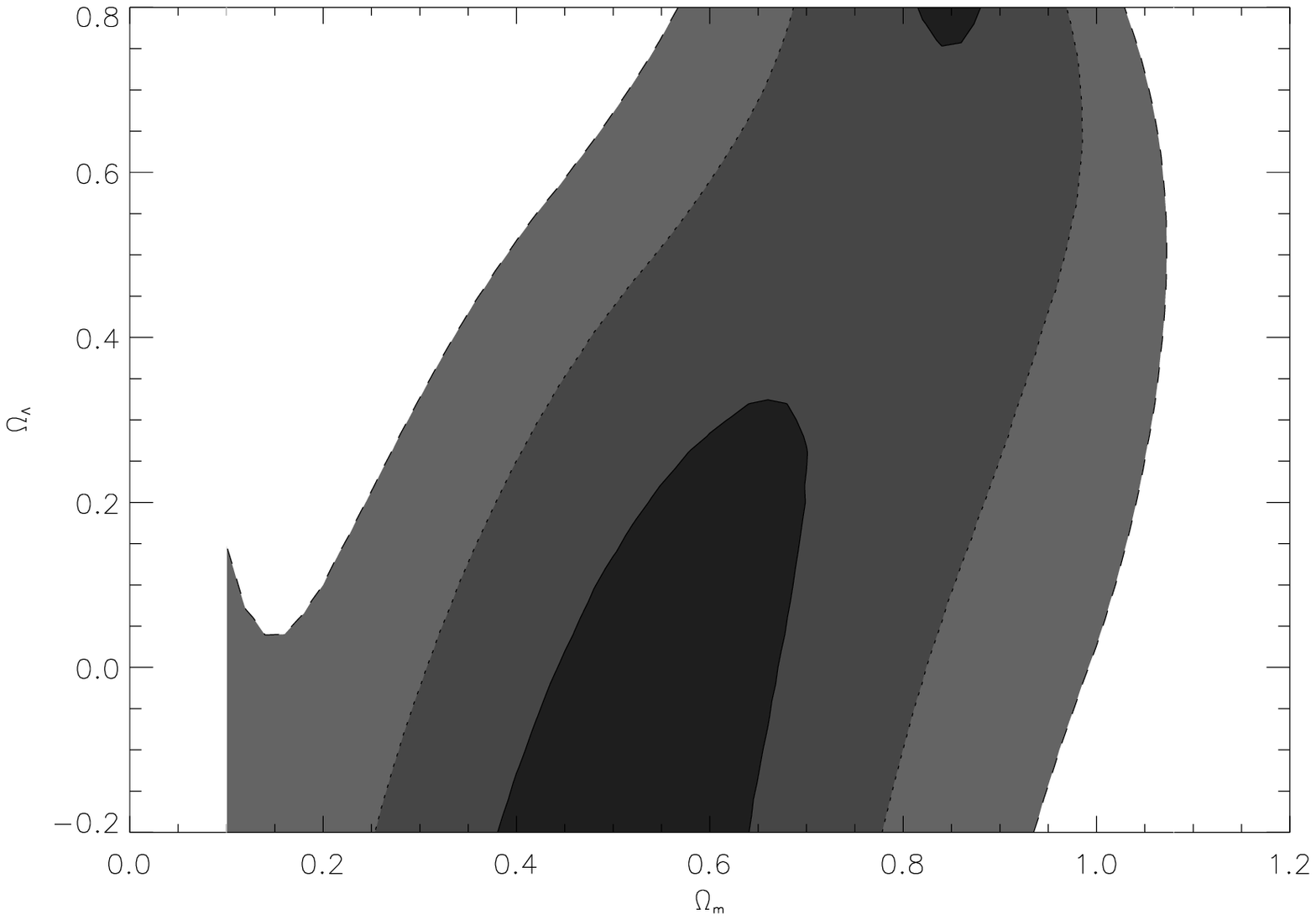}
\caption[]{The confidence contours for $\Omega_m$ and $\Omega_{\Lambda}$ at the 1, 2, and 3 $\sigma$ level on 1 parameter using the cosmological test based on the gas fraction of distant and nearby clusters and assuming  $A_{TM}=6.24$ keV. \emph{Top}: Test performed at R$_{2}$. \emph{Bottom}: Test performed at the virial radius.}
\end{figure}
We can point out that, as expected, no difference was found in the results at the radii studied by S05 (R$_{vir}$, R$_{1}$, and R$_{2}$) when using the two alternative ways to determine the virial radius and mass (comparing to the mean or the critical density of the Universe). We have also determined the confidence contours in the two-dimensional parameter space of $\Omega_m - \Omega_{\Lambda}$. For this, we used the formulae provided by \cite{Pierpaoli} to calculate the virial overdensity $\Delta_V$ for any set of cosmological parameters and followed the same analysis described above.
Results are presented in Fig. 2 for the test at the virial radius and at R$_{2}$. The contours are compatible with the likelihood functions obtained in the case of a flat Universe and they clearly show the significant difference in the results of the gas fraction test when we consider an inner or outer radius to perform it. 

\section{Discussion}

We have used combined data from \emph{Chandra} and \emph{XMM-Newton} to apply the cosmological test based on the gas fraction evolution considering that this quantity varies with the temperature of the cluster and that this dependence should evolve with redshift through a simple scaling argument based on the gravitational growth of structure in an expanding Universe. The obtained results allow us to conclude that this kind of test is highly dependent on the radius considered to implement it, since we found significantly different values for $\Omega_m$ when the test is performed at different normalized radii. The complexities in cluster's internal structures which were evidenced in S05 are at the origin of this observed ambiguity. It seems that the gas fraction observed in low and high-redshift clusters cannot be accounted for by a purely gravitational heating of the gas, which in other words means that non-gravitational effects present in clusters induce a significant deviation of the gas fraction profile from standard scaling. We found no significant differences in the results when using two different definitions for normalizing the radius, and it is interesting to point out that the results obtained at R$_{2500}$ are compatible with those obtained by Allen et al.(2002) and Allen et al.(2004) at the same radius. Note although that these authors did not apply the same method, as they assumed a constant gas fraction at their working radius. 

The main question raised by this work is whether there is a radius at which scaling relations can be used with more confidence. At first one is tempted to say that the virial radius should provide a positive answer, since at this radius non-gravitational processes are thought to be less important. However, there are two aspects that do not allow us to make definite conclusions on this point: a) the observations of high-redshift clusters do not extend up to the virial radius and so the gas masses at this region were obtained by extrapolating the $\beta$-model fitted on the observed parts of the clusters, and b) the maximum likelihood values for $\Omega_m$ at this radius are too high ($\Omega_m \sim 0.7$), which translates to a significant deviation from the concordance model supported by other observational data such as CMB, Type Ia supernovae, and baryonic oscillations (\cite{Spergel}, \cite{astier}, \cite{eisenstein}, \cite{Blanchard}). Observations of X-ray emissivity in distant clusters up to the virial radius will be critical to providing definitive answers on the first aspect. Despite this, the use of gas mass fraction evolution (or lack of it) in clusters as a precision method for constraining cosmological parameters seems to be seriously compromised.

\begin{acknowledgements}
This research has made use of the X-ray cluster database (BAX), which is operated by the Laboratoire Astrophysique de Toulouse-Tarbes (LATT), under contract with the Centre National d'Etudes Spaciales (CNES). L. Ferramacho acknowleges financial support provided by Funda\c c\~ao para a Ci\^encia e Tecnologia (FCT, Portugal) under fellowship contract SFRH/BD/16416/2004.  

\end{acknowledgements}


\begin{thebibliography}{99}
\bibitem[Allen et al. 2002]{Allen2002} Allen, S. W., Schmidt, R. W., \& Fabian, A. C. 2002, MNRAS, 334, L11
\bibitem[Allen et al. 2004]{Allen2004} Allen, S.W, Schmidt, R.W, Ebeling, H, Fabian, van Speybroeck, L. 
\bibitem[(Arnaud et al. 2002)]{Arnaud} Arnaud, M., 
Majerowicz,  S., Lumb, D. H, Neumann,  D. M.,  Aghanim N.,  Blanchard, A., Boer,
M., Burke, D., Collins, C.,  Giard, M.,  Nevalainen, J., Nichol,  R. C., 
 Romer, K. \& Sadat, R. 2002,  {  A\&A} 390,  27
\bibitem[Astier et al. 2006]{astier} Astier, P., Guy, J., Regnault, N., Pain, R. et al. 2006, {A\&A} 447, 31
\bibitem[(Bartlett et al. 2001)]{bartlett} Bartlett, J. G.  et al., 2001, proceedings of the XXI rencontres de Moriond, astro-ph/0106098
\bibitem[Blanchard \& Douspis (2005)]{blancharddouspis}Blanchard, A. \& Douspis. M. 2005, {  A\&A} 436, 411
\bibitem[Blanchard et al. 2006]{Blanchard}Blanchard, A., Doupis, M., Rowan-Robinson, M., Sarkar, S. 2006, A\&A, 449, 925
\bibitem[Bryan \& Norman (1998)]{BN}Bryan, G.L. \& Norman, M.L. 1998, {  ApJ} 495, 80
\bibitem[(Cavaliere \& Fusco-Femiano 1976)]{Cavaliere}Cavaliere, A. \& Fusco-Fermiano, R. 1976, {A\&A}, 49, 137
\bibitem[Eisenstein et al. 2005]{eisenstein} Eisentein, D. J., Zehavi, I., Hogg, D. W., et al. 2005, {ApJ} 633, 560
\bibitem[Ettori \& Fabian 2000]{ettori1}Ettori, S., Fabian, A. 2000, ASPC 200, 369
\bibitem[Ettori et al. 2003]{Ettori}Ettori, S., Tozzi, P., Rosati, P. 2003, {A\&A} 398, 879 
\bibitem[Evrard, Metzler \& Navarro 1996]{evrard} Evrard, A. E., Metzler, C. A., \& Navarro, J. F., 1996, {ApJ} 469, 494 
\bibitem[Finoguenov et al. 2001]{Finoguenov}Finoguenov, A., Reiprich, T. H. \& Bohringer, H. 2001, {  A\&A} 368, 749
\bibitem[Horner et al. 1999]{Horner}Horner, D. J., Mushotzky, R. F. \& Scharf, C. A. 1999, {  ApJ} 520, 78 
\bibitem[Keiser 1986]{Keiser}Kaiser, N. 1986, MNRAS 222, 323
\bibitem[Lumb et al. (2004)]{lumb} Lumb, D. H.,  Bartlett,  J. G.,  Romer, K.,
Blanchard, A.,  Burke, D., Collins, C., Nichol,  R.C.,  Giard, M.,  Marty, P., Nevalainen, J.,  Sadat, R. \& Vauclair, S. 2004, A\&A 420, 853 
\bibitem[(Navarro, Frenk \& White 1997 )]{nfw} Navarro, J. F., Frenk, C. S. \& White, S. D. M. 1997, {  ApJ} 490, 493
\bibitem[Nevalainen et al. 2000]{Nevalainen}Nevalainen, J., Markevitch, M. \& Forman, W. 2000, {ApJ} 532, 694
\bibitem[Pedersen \& Dahle 2006]{Pedersen} Pedersen, K., Dahle, H., 2006, astro-ph/0603260
\bibitem[Pen (1997)]{pen}Pen, Ue-Li 1997, NewA, 2, 309
\bibitem[Pierpaoli et al. (2001)]{Pierpaoli}Pierpaoli, E., Scott, D. \& White, M. 2001, MNRAS 325, 77
\bibitem[Rines et al. 1999]{Rines}Rines, K., Forman, W., Pen, U., Jones, C. 1999, {ApJ} 517, 70 
\bibitem[Sadat et al. (2005)]{sadat05} Sadat, R., Blanchard, A., Vauclair, S. C. et al. 2005, {A\&A}, 437, 310
\bibitem[Sanderson et al. 2003]{Sanderson}Sanderson, A. J. R., Ponman, T. J., Finoguenov, A., Lloyd-Davies, E. J. \& Markevitch, M. 2003, {  MNRAS} 340, 989
\bibitem[Sasaki (1996)]{sasaki}Sasaki, S. 1996, {  PASJ} 48, 119
\bibitem[Spergel et al. 2003]{Spergel}Spergel, D.N., Verde, L., Peiris, H.V., Komatsu, E., Nolta, M.R., et al. 2003, ApJS, 148, 175 
\bibitem[Vikhlinin et al. (1999)]{vfj99}{Vikhlinin}, A., {Forman}, W. R. \& {Jones}, C. 1999 , {  ApJ} 525, 47
\bibitem[Vikhlinin et al. (2002)]{vik2002} {Vikhlinin}, A.,  Van Speybroeck, Markevitch, M., Forman, W. R. \& Grego, L. 2002, {  ApJ} 578, L107






\end{thebibliography}
\end{document}